\documentstyle[aas2pp4]{article}
\def \hb {H$\beta$}
\def \oIII {[O~III]~$\lambda 5007$}
\def \kms {~km~s$^{-1}$}
\begin{document}

\title{A NOTE ON THE VIABILITY OF GASEOUS IONIZATION IN ACTIVE GALAXIES
BY FAST SHOCKS}
\author{Ari Laor}
\affil{Physics Department, Technion, Haifa 32000, Israel\\
laor@physics.technion.ac.il}

\begin{abstract}
Morphological and spectroscopic evidence suggest that shocks may affect the 
spatial and velocity distributions of gas in the narrow line region (NLR) 
and extended NLR of some active galaxies. It thus seemed plausible that
shocks may also energize the NLR.
The observed emission line ratios strongly favor photoionization as the
heating source for this gas, but it is not clear whether the 
ionizing radiation is generated in the NLR 
by ``photoionizing shocks'' or whether
the ionizing radiation originates at the central continuum source. Here I 
point out that shocks are highly inefficient in producing line emission. 
Shocks in the NLR can convert at most
$\sim 10^{-6}$  of the rest mass to ionizing radiation, compared with 
a maximum conversion efficiency
of $\sim 10^{-1}$ for the central continuum source. 
The required mass flow rate through ``photoionizing shocks'' in the NLR is 
thus a few orders of magnitude higher than the mass accretion rate
required to power the NLR by the central continuum source. 
Since gravity appears to dominate the 
NLR cloud dynamics, shocks must lead to an inflow, and 
the implied high inflow rates can be ruled out in most active galaxies.

NLR dynamics driven by a thermal wind or by some jet configurations
may produce the mass flux through shocks required for photoionizing shocks
to be viable, but the mass flux inward from the NLR must be kept
$\sim 100$-1000 times 
smaller.

Photoionizing shocks are a viable mechanism in very low luminosity
active galaxies if they are highly sub-Eddington ($\lesssim 10^{-4}$) 
and if they convert mass to radiation with a very low efficiency 
($\lesssim 10^{-4}$).

\end{abstract}

\keywords{galaxies: active-galaxies: Seyfert- shock waves}

\section{INTRODUCTION}
Active Galactic Nuclei (AGNs) display emission lines originating from clouds 
in the narrow line region (NLR) and the broad lines region 
(BLR; e.g. Netzer 1990). 
The line emission could be powered either locally through 
cloud-cloud 
collisions (Daltabuit, MacAlpine \& Cox, 1978), or by absorption of 
ionizing radiation from the continuum source at the center 
(e.g. Davidson \& Netzer
1979). The time lagged response of the broad emission lines to UV continuum 
variability (e.g. Peterson 1993) provides nearly conclusive evidence that 
the BLR clouds are powered by the central continuum source. The same 
relation cannot be established for the NLR since reverberation
timescales are far too long. The observed line ratios in the NLR 
argue strongly in favor of pure photoionization 
(e.g. Ferland \& Osterbrock 1986), which suggests that the NLR is 
also powered by the central continuum source. However, 
some models suggest that shocks due to cloud-cloud collisions may provide 
significant heating for the NLR, in addition to the central continuum
source (e.g. Contini \& Aldrovandi 1983, Contini 1997). 

Renewed interest in shocks as a viable energy source came following the
detailed calculations of Dopita \& Sutherland (1995, 1996) who
have shown that fast shocks produce strong ionizing radiation
which can be reprocessed into line emission.
Thus, although the emission line spectrum from the NLR strongly suggests
that clouds are powered by photoionization, rather than collisional 
ionization, this power could be produced by shocks in the NLR, rather
than by the central continuum source. 
Morphological and spectroscopic evidence suggest that jet-cloud interactions
may produce such shocks in the extended NLR of nearby
AGNs (e.g. Cecil, Morse, \& Veilleux, 1995; 
Capetti, Axon, \& Macchetto 1997; Falcke, Wilson \& Simpson 1998). 
For a comprehensive review and 
a critical assessment of the 
available observational evidence for photoionizing shocks in AGNs
see Morse, Raymond \& Wilson (1996) and Wilson (1997). 

The purpose of this note is to point out that shocks are highly
inefficient, compared with the central AGN, in converting mass to
ionizing radiation, and can thus be ruled out
as a significant source of energy for the NLR for most AGNs.

\section{QUALITATIVE ESTIMATES}

Shocks convert part of the directed kinetic energy of the colliding 
gas clouds to thermal energy.
If all the kinetic energy of cloud-cloud collision in the NLR is converted to
heat, all the heat is converted to ionizing radiation, and all the
ionizing radiation is absorbed and converted to line emission, then
the maximum fraction of rest mass converted to line emission would
be $\epsilon_{\rm shock}=v^2/2c^2$. The typical velocity of gas in the 
NLR is 
$\sim 250-500$~\kms (see \S 3), and thus the maximum efficiency of rest mass to
line emission conversion is $\epsilon_{\rm shock}\sim 10^{-6}$. 

If the gas reaches the center of the active nucleus, about 10\% of its 
rest mass is converted
to heat, which may be radiated away and converted to line emission, yielding
a maximum efficiency of $\epsilon_{\rm AGN}\sim 0.1$.
Thus, basic energy conservation implies that the photoionizing shocks 
scenario requires a mass flow through NLR shocks which is a few
orders of magnitude larger than required into the 
central photoionizing source. A more
accurate estimate of the required accretion rates in both scenarios
is given below.

\section{QUANTITATIVE ESTIMATES}
The \oIII\ line is typically the strongest narrow line in bright
active galaxies
(e.g. Ferland \& Osterbrock 1986; Veilleux \& Osterbrock 1987), and I
therefore use it below as a measure of the NLR luminosity.

Dopita \& Sutherland (1995) derived the following scaling relation for
the \hb\ flux per unit area generated in the cooling gas behind a
fast photoionizing shock
\[ 
l_{{\rm H}\beta}=
7.44\times 10^{-6}v_{100}^{2.41}n\ \ \ {\rm ergs~cm}^{-2}~{\rm s}^{-1}\ , 
\]
where $v=100v_{100}$~\kms\ is the shock velocity and $n$ is the  
ambient gas density. Additional line emission is produced in the
unperturbed gas in front of the shock, the "precursor H~II region",
which is given by
\[ 
l_{{\rm H}\beta}=
9.85\times 10^{-6}v_{100}^{2.28}n\ \ \ {\rm ergs~cm}^{-2}~{\rm s}^{-1}\ . 
\]
The highest \oIII/\hb\ flux ratio found by Dopita \& Sutherland is
$\sim 13$, and we therefore assume 
$l_{\rm [O~III]}\le 13 l_{{\rm H}\beta}$.
 The mass flux through the shocks is 
\[ 
{\mathaccent 95 M}_{\rm shock}=A_snv, 
\]
where $A_s=L_{\rm [O~III]}/l_{\rm [O~III]}$ is 
the shock area, and $L_{\rm [O~III]}$ is the intrinsic \oIII\
luminosity. 
 Thus, using the shock + precursor line emission, one gets
\[
{\mathaccent 95 M}_{\rm shock}\ge 2730 L_{42} v_{100}^{-1.41}(1+1.32v_{100}^{-0.13})^{-1}
\ \ M_{\odot}~{\rm yr}^{-1}\ , 
\]
where $L_{\rm [O~III]}= 10^{42}L_{42}$~erg~s$^{-1}$.

The relation between the mass accreted by the AGN, 
${\mathaccent 95 M}_{\rm AGN}$, and $L_{\rm [O~III]}$ resulting from photoionization
by the AGN is not unique. This relation depends on the accretion efficiency,
the covering factor of the NLR gas, the ionization parameter 
and density of
the NLR gas, and the presence of dust within the NLR gas. Rather than
deriving a theoretical relation between ${\mathaccent 95 M}_{\rm AGN}$ and 
$L_{\rm [O~III]}$,
which would require knowledge of the parameters mentioned above, I use 
 the observed
 $L_{\rm [O~III]}$ versus ${\mathaccent 95 M}_{\rm AGN}$ relation. 
The continuum $\nu L_{\nu}$ at 5000~\AA\ is given by 
$L_{\rm [O~III]}\times 5000/EW_{\rm [O~III]}$,
where $EW_{\rm [O~III]}$ is the equivalent width of [O III] 
in units of \AA. 
The bolometric luminosity is
$L_{\rm Bol} \sim 10\times \nu L_{\nu}(5000~{\rm \AA})$ 
(e.g. fig.7 in Laor \& Draine
1993), and the accretion rate is 
${\mathaccent 95 M}_{\rm AGN}=L_{\rm Bol}/\epsilon_{\rm AGN} c^2$. These
relations give
\[ {\mathaccent 95 M}_{\rm AGN}=0.293 L_{42} EW_{30}^{-1}\epsilon_{0.1}^{-1}\ \ 
M_{\odot}~{\rm yr}^{-1}\ , \]
where $EW_{\rm [O~III]}=30 EW_{30}$~\AA, and 
$\epsilon_{\rm AGN}=0.1\epsilon_{0.1}$.

The ratio ${\mathaccent 95 M}_{\rm shock}/{\mathaccent 95 M}_{\rm AGN}$ is a function of
$v_{100}, EW_{30}$ and $\epsilon_{0.1}$. Below we estimate
typical values and dispersions for these parameters. 

The $[O~III]$ FWHM data for 71 broad line active galaxies presented 
in Fig.6a of Whittle (1985) gives 
$\langle v_{100}\rangle =3.5^{+1.4}_{-1}$, where the quoted range
includes 2/3 of the objects (equivalent to $\pm 1 \sigma$ for
a normal distribution). Data on the $EW_{\rm [O~III]}$ of 36 Seyfert 1 
galaxies in Osterbrock (1977) gives 
$\langle EW_{\rm [O~III]}\rangle =39\pm 22$, and 
the complete sample of Boroson \& Green (1992) gives
$\langle EW_{\rm [O~III]}\rangle =37^{+40}_{-23}$ for 18 Seyfert 1 galaxies and
$\langle EW_{\rm [O~III]}\rangle =20^{+22}_{-14}$ for 69 quasars.
We therefore adopt $v_{100}\sim 3.5$, and $EW_{30}\sim 1$ as typical 
values, which gives
\[ \frac{{\mathaccent 95 M}_{\rm shock}}{{\mathaccent 95 M}_{\rm AGN}}\sim 750\epsilon_{0.1}, \]
with a typical range of about a factor of two above and below.

The value of $\epsilon_{0.1}$ can be estimated by comparing the time 
integrated density of quasar light 
with the current density of massive black holes, as determined by the
black hole to bulge mass correlation suggested by Magorrian et al. (1997).
This comparison
provides a rather strict lower limit on AGNs time average accretion
efficiency of $\epsilon_{0.1}\simeq 0.09$
(see equations 1 \& 2 in Haehnelt, Natarajan, \& Rees 1997 with
$h=0.7$ and $f_B=0.1$). Larger values for the mean observed
$\epsilon_{0.1}$ are obtained if
not all bulges contain massive black holes, or if some of the black holes
growth occurs in an unobserved phase with $\epsilon_{0.1}\ll 1$ 
(see discussion in Haehnelt et al.).
Typical accretion scenarios give $\epsilon_{0.1}\sim 1$. 
Thus, ${\mathaccent 95 M}_{\rm shock}$ is at least about two,
and more likely nearly three orders of 
magnitude larger than ${\mathaccent 95 M}_{\rm AGN}$.

\section{DISCUSSION}

\subsection{Gravity Powered Shocks}

If the gas velocity in the NLR is dominated by gravity, as strongly
suggested by observations (Whittle 1992a, 1992b), then the loss of
kinetic energy in shocks necessarily implies a loss of angular momentum, 
which must lead to an inflow. If there is no accumulation of mass 
somewhere between the central
black hole and the NLR then ${\mathaccent 95 M}_{\rm shock}$ eventually becomes
${\mathaccent 95 M}_{\rm AGN}$. Having ${\mathaccent 95 M}_{\rm shock}={\mathaccent 95 M}_{\rm AGN}$ 
requires $\epsilon_{\rm 0.1}\sim 10^{-3}$. 
However, the estimates
of Haehnelt et al. mentioned above provide a lower limit of 
$\epsilon_{\rm 0.1}\simeq 0.1$ for the average efficiency. 
Thus, if the NLR of most AGNs is powered by shocks, 
then the mass flow through the NLR cannot accumulate in the central black hole. 
 Can ${\mathaccent 95 M}_{\rm shock}$ accumulate somewhere between the central 
black hole and the NLR?

The black hole mass estimates of Magorrian et al. and of earlier
studies of the cores of normal galaxies (e.g. 
Kormendy \& Richstone 1995) are based on 
stellar velocity dispersions in the central 10-100 pc. Thus, the
limits on the accretion efficiency apply on the scale of the NLR
as well. Stellar velocity dispersions
in a few nearby AGNs also indicate that the mass interior to 
the NLR is significantly smaller than expected from
shock excitation (Oliva 1997). Thus, as pointed out
by Oliva (1997) for the coronal line region in NGC 1068, there is no 
place to hide the large ${\mathaccent 95 M}_{\rm shock}$ implied by the photoionizing 
shocks scenario for an astrophysically interesting period of time.

Cloud-cloud collisions as a source of photoionizing shocks are a 
viable mechanism only in low luminosity
AGNs where $L/L_E\lesssim 10^{-4}$, such as the Galactic center 
(Narayan, Yi, \& Mahadevan 1995), M87 (Reynolds et al. 1996, see
a photoionizing shock model in Dopita et al. 1997),
and possibly LINERS (Ho, Filippenko, \& Sargent 1993),
since such objects can maintain $\epsilon_{\rm 0.1}\lesssim 10^{-3}$,
without accumulating excessive mass, 
for a non negligible fraction of their lifetime. The observed emission
spectrum of LINERS, however, appears to argue against photoionizing shocks 
(Ferland \& Netzer 1983; Filippenko 1985; Ho, Filippenko, \& Sargent
1993, 1996; Maoz et al. 1995).

\subsection{Wind Powered Shocks}

Shocks in the NLR would not necessarily lead to an
inflow if the gas motion is not dominated by gravity. 
In particular, the shocks may be produced by 
interaction of NLR clouds with an outflowing wind. The clouds cannot
be in pure radial outflow since that would result in mass ejection
at a rate ${\mathaccent 95 M}_{\rm shock}$ and 
mass and energy conservation requires a time average
${\mathaccent 95 M}_{\rm shock}<{\mathaccent 95 M}_{\rm AGN}$, which can be maintained only when
$\epsilon_{0.1}\lesssim 10^{-3}$. One may imagine instead a ``turbulent''
velocity field where the required 
${\mathaccent 95 M}_{\rm shock}\sim 220L_{42}M_{\odot}~{\rm yr}^{-1}$ (for
$v_{100}=3.5$) flows in and
out within the NLR, with only 0.1-1\% of this accretion rate being
able to flow further inward to provide  ${\mathaccent 95 M}_{\rm AGN}$.

What can power such a wind?  One possibility may be a radiation-pressure
driven wind (e.g. on dust grains). The maximum momentum flux available in
this case is ${\mathaccent 95 P}_{\rm rad}=L_{\rm Bol}/c$, or 
${\mathaccent 95 P}_{\rm rad}={\mathaccent 95 M}_{\rm AGN}
\epsilon_{\rm AGN}c$. 
The momentum flux
in shocked gas is ${\mathaccent 95 P}_{\rm shock}=
{\mathaccent 95 M}_{\rm shock}v$, and the maximum
ratio between available and required momentum fluxes,   
\[ {\mathaccent 95 P}_{\rm rad}/{\mathaccent 95 P}_{\rm shock}=
0.1\epsilon_{0.1}c/750\epsilon_{0.1}v, \] 
is about an order of magnitude too small.

Another possibility is a thermal-pressure driven wind, in which case
momentun is not conserved locally, but only globaly. 
What can heat this wind? The wind velocity
should be of the order of 250-500~km~s$^{-1}$ which corresponds to
$T\sim 1-3\times 10^7$~K. This temperature range is obtained naturally
for Compton-heated winds in AGNs (Begelman, McKee, \& Shields 1983; Mathews
\& Ferland 1987). The photoionizing energy flux in the NLR
is $E_{\rm ph}n_{\rm ph}c$, where $E_{\rm ph}\sim 30$~eV is the mean
photon energy, and $n_{\rm ph}$ is the photon density. The mechanical energy flux
provided by the wind is $E_{\rm p}2nc$, where $E_{\rm p}\sim 1$~keV 
is the mean
particle energy. The photoionizing/mechanical flux ratio is thus $13U$, where
$U\equiv n_{\rm ph}/n$ is the wind ionization parameter. The Compton temperature
is obtained for $U>100$, and thus mechanical (i.e. shock)
 heating would be a negligible effect
compared with direct photoionization heating.

Alternatively, a thermal wind may be mechanically heated by the dissipation of
a jet kinetic energy, as seen for example in jet cocoons. 
The rate of energy deposition by the wind into the shocks 
is roughly $\frac{3}{16}{\mathaccent 95 M}_{\rm shock}v^2=3.2\times 10^{42}L_{42}$~erg~s$^{-1}$
(for $v_{100}=3.5$),
which should be provided by the jet. Since 
$L_{\rm Bol}= 1.7\times 10^{45} L_{42} EW_{30}^{-1}$, the jet requires 
only $\sim 0.2$\% of $L_{\rm Bol}$.
The energy deposited in the shocked gas should alos be observable as an 
extended non-variable source of continuum emission in the optical to 
soft X-ray range. Additional spatially extended free-free emission at 
$T\sim 1-3\times 10^7$~K should come from the thermal wind, but the luminosity
of this component could be well below $10^{42}$~erg~s$^{-1}$.

\subsection{Jet Powered Shocks}

Although gravity appears to play the key role in the NLR, Whittle (1992a, 1992b) 
finds evidence for excess velocity dispersions in a small fraction of AGNs
(those with a high radio luminosity and linear radio morphology), which can be 
attributed to a jet-cloud interaction (see also Bicknell et al. 1997). 
Thus one can imagine a picture where clouds are swept up by the jet, and later
fall back in, with nearly zero net accretion rate through the NLR, just as in 
the wind powered shocks picture.

The jet starts at the center with velocity  $v_{\rm jet}$ and mass flux
${\mathaccent 95 M}_{\rm jet}$, gradually entraining local gas until its
mass flux increases to ${\mathaccent 95 M}_{\rm shock}$ and its velocity drops to
$v$. 
Mass, momentum, and energy conservation are used below to constrain the 
properties of such a jet. The jet motion is assumed to be balistic
and any possible interaction with a confining
medium is neglected.

Momentum flux conservation gives
\[
{\mathaccent 95 M}_{\rm jet}=
v{\mathaccent 95 M}_{\rm shock}/\gamma_{\rm jet}\beta_{\rm jet}c\ ,
\]
where $\beta_{\rm jet}=v_{\rm jet}/c$ and 
$\gamma_{\rm jet}=(1-\beta_{\rm jet}^2)^{-1/2}$ 
(note that $\beta_{\rm shock}\lesssim 0.0017$,  
$\gamma_{\rm shock}\sim 1$). Mass flux conservation requires
${\mathaccent 95 M}_{\rm jet}<{\mathaccent 95 M}_{\rm AGN}$ which together with the above expression
gives
\[
\gamma_{\rm jet}\beta_{\rm jet}c/v>{\mathaccent 95 M}_{\rm shock}/{\mathaccent 95 M}_{\rm AGN}\ .
\]
Using the ${\mathaccent 95 M}_{\rm shock}/{\mathaccent 95 M}_{\rm AGN}$ relation obtained in \S 3 
(note that this relation is independent of the presence of a jet),
and $v=350$~km~s$^{-1}$ gives 
$\gamma_{\rm jet}\beta_{\rm jet}>0.875\epsilon_{0.1}$ or
\[ 
\beta_{\rm jet}>0.875 \epsilon_{0.1}/\sqrt{1+0.77\epsilon_{0.1}^2}\ .
\]
This is the minimum jet velocity which can provide the required
momentum flux without exceeding the mass flux available through accretion.

The flux of kinetic energy in the jet must be smaller than the energy
flux generated by the accretion,
i.e.
\[
(\gamma_{\rm jet}-1){\mathaccent 95 M}_{\rm jet}c^2<0.1\epsilon_{0.1}{\mathaccent 95 M}_{\rm AGN}c^2\ .
\]
Substituting for ${\mathaccent 95 M}_{\rm jet}$ from the momentum flux conservation, and
using the ${\mathaccent 95 M}_{\rm shock}/{\mathaccent 95 M}_{\rm AGN}$ relation obtained in \S 3
gives
\[
0.1\epsilon_{0.1}\gamma_{\rm jet}\beta_{\rm jet}c/
(\gamma_{\rm jet}-1)v>750 \epsilon_{0.1} \ ,
\]
which for $v=350$~km~s$^{-1}$ gives
\[
\beta_{\rm jet}<0.26\ .
\]
This is the maximum jet velocity which can provide the required
momentum flux without exceeding the energy flux available through accretion.
The upper and lower limits on $\beta_{\rm jet}$ obtained above 
allow a solution for $\beta_{\rm jet}$ only if $\epsilon_{0.1}<0.26$.
Thus, for jet entrainment to be viable the jet must be subrelativistic,
and the accretion efficiency must be fairly low.

As the jet entrains ambient gas it slows down, while conserving
momentum and losing kinetic energy.
The lost kinetic energy is converted to internal energy of
the entrained gas and some of it will be radiated away.
 For example, in a jet with a solid angle of 
$\Omega/4\pi=0.01$, reaching a final $v_{100}=3.5$, i.e. 
${\mathaccent 95 M}_{\rm shock}=220
L_{42} M_{\odot}~{\rm yr}^{-1}$, at $r=10$~pc, and $T=10^7$K, the
free-free cooling time is $\sim 6$~yr, which is well below the dynamical 
time scale, $r/v\sim 3000$~yr. The jet cooling will produce an  
extended continuum source, in addition to that within the shocked gas
itself (\S 4.2). This continuum can be quite intense. The initial
jet kinetic energy is 
$L_{\rm jet}^{\rm in}\simeq \frac{\beta_{\rm jet}}{0.26} L_{\rm Bol}$,
the accretion efficiency is $\epsilon_{0.1}>0.1$ (\S 4.1), which gives
a minimum jet
velocity of $\beta_{\rm jet}\simeq 0.087$ (from momentum conservation).
Since the final jet kinetic energy is only
$L_{\rm jet}^{\rm f}\simeq 0.005 L_{\rm Bol}$, the dissipated
kinetic energy in the entrained gas is large $\sim 0.3L_{\rm Bol}$.
A significant fraction of this internal energy is likely to be radiated    
through thermal and nonthermal continuum
emission somewhere in the X-ray to radio bands. Thus, observational
constraints on the amount of spatially extended continuum emission
in nearby AGNs can be used to constrain the feasibility of the jet-cloud
interaction scenario. Wilson, Ward \&
Haniff (1987) find that the extended radio power in nearby AGNs is 
$\le 10^{-2}L_{\rm NLR}$. Comparable constraints in other bands would
rule out jet entrained gas as the source of power for the NLR in most AGNs.

If the jet is still confined as it entrains ambient gas,
if cooling is negligible, and if the jet cross section increases outward, 
then the random particle motion can be converted back to directed motion.
The jet basically converts kinetic energy to internal energy and back to
kinetic energy, conserving kinetic energy and not conserving momentum
(due to the confining medium).
The limits obtained above will not be valid (see Phinney 1983 for a 
comprehensive discussion), and the jet just
needs to provide the kinetic energy deposited in the shocks 
($0.002L_{\rm Bol}$), as in the
case of the thermal pressure driven wind discussed above.
 
A jet may also power the NLR by directly dumping its kinetic
energy into gas clouds in the NLR, rather than by inducing gas 
motion which results in photoionizing shocks, as assumed above.
However, in this case the line emission would result from
thermal excitation, rather than photoionization, which is not consistent 
with the observed spectrum from the NLR.

\section{CONCLUSIONS}

Photoionizing shocks due to cloud-cloud collisions in a gravity dominated NLR
cannot provide the source of energy for narrow line emission in
most AGNs. 

NLR dynamics dominated by forces other than gravity may allow the high
mass flux through shocks required for photoionizing shocks
to be viable. The clouds velocities can be produced by a 
wind-cloud interaction, which
requires a mechanically heated thermal wind. 
Alternatively, the clouds velocities may be dominated by a jet-cloud
interaction. This either implies a strong extended continuum source, which
is not yet observed, or otherwise specific jet configurations. In any
non-gravitational cloud dynamics picture the mass flux inward of the 
NLR must be kept $\sim 100$-1000 times smaller than through the NLR shocks.

Photoionizing shocks could be relevant in very low luminosity
AGNs, such as LINERs, if these objects are both highly sub Eddington
($L/L_E\lesssim 10^{-4}$) and if they convert mass to radiation with a 
very low efficiency ($\epsilon_{\rm AGN}\lesssim 10^{-4}$).  

\acknowledgments

Thoughtful and constructive comments by Sterl Phinney, Dani Maoz, and the referee 
are greatly appreciated. 
This research was supported in part by a grant from the Israel Science Foundation,
by the E. and J. Bishop research fund, and by the Milton and Lillian Edwards 
academic lectureship fund. 

%\newpage

\end{document}